\documentclass[10pt]{article}
\usepackage{graphicx}
\usepackage{amsmath}
\usepackage{amssymb}
\usepackage{caption2}
\begin{document}
\bibliographystyle{prsty}
\begin{center}
{\large {\bf \sc{  Analysis of the $\Lambda_Q$  baryons in the nuclear matter  with   the   QCD sum rules  }}} \\[2mm]
Zhi-Gang Wang \footnote{E-mail,wangzgyiti@yahoo.com.cn.  }   \\
 Department of Physics, North China Electric Power University, Baoding 071003, P. R. China
\end{center}

\begin{abstract}
In this article, we study   the $\Lambda_c$ and $\Lambda_b$  baryons in the nuclear matter using the QCD sum rules, and obtain the in-medium  masses   $M_{\Lambda_c}^*=2.335\,\rm{GeV}$,
 $M_{\Lambda_b}^*=5.678\,\rm{GeV}$, the in-medium vector self-energies $\Sigma^{\Lambda_c}_v=34\,\rm{MeV}$,  $\Sigma^{\Lambda_b}_v=32\,\rm{MeV}$, and
  the in-medium pole residues
 $\lambda_{\Lambda_c}^*=0.021\,\rm{GeV}^3$,
 $\lambda_{\Lambda_b}^*=0.026\,\rm{GeV}^3$.  The mass-shifts are
 $M_{\Lambda_c}^*-M_{\Lambda_c}=51\,\rm{MeV}$ and
 $M_{\Lambda_b}^*-M_{\Lambda_b}=60\,\rm{MeV}$, respectively.
\end{abstract}



\section{Introduction}
In the past years, there have been several  progresses
on the spectroscopy of the heavy baryon states.  Several new  charmed baryons, such as the $\Omega_{c}^{*}(2770)$,
$\Lambda_c^+(2765)$, $\Lambda_c^+(2880)$, $\Lambda_c^+(2940)$,
$\Sigma_c^+(2800)$, $\Xi_c^+(2980)$, $\Xi_c^+(3080)$,
$\Xi_c^0(2980)$, $\Xi_c^0(3080)$,   have been observed  in recent years
\cite{ShortRV}, and re-vivified  the interest in the
  spectroscopy of the charmed  baryons. On the other hand, the
  in-medium properties of hadrons play an important role  in
understanding the strong interactions, the relativistic  heavy ion collisions and
the nuclear astrophysics.  The upcoming  FAIR
 project at GSI
 provides the opportunity to extend the experimental
studies of the in-medium hadron properties into the charm sector. The CBM
collaboration  intends to study the in-medium properties of the
 hadrons \cite{CBM},
 while the $\rm{\bar{P}ANDA}$ collaboration will focus on  the charm spectroscopy,
 and mass and width modifications  of the charmed hadrons
 in the nuclear matter \cite{PANDA}.
  So it is interesting to study the in-medium properties of the charmed baryons.

  The  QCD sum rules is a powerful theoretical tool in studying the
ground state hadrons \cite{SVZ79}. In the
QCD sum rules, the operator product expansion is used to expand the
time-ordered currents into a series of quark and gluon condensates
which parameterize the long distance properties. Based on
the quark-hadron duality, we can obtain copious information about
the hadronic parameters at the phenomenological side
\cite{SVZ79}. The in-medium properties of the light flavor hadrons have been studied extensively with the
QCD sum rules \cite{Drukarev1991,C-parameter}, while the works on the in-medium properties of the
 heavy hadrons   focus on the $J/\psi$, $\eta_c$, $D$,  $B$, $D_0$ and $B_0$ \cite{Jpsi-etac,Hayashigaki}.
  In this article, we study the in-medium properties of the $\Lambda_c$ baryon in the nuclear
matter using the QCD sum rules, furthermore, we study
the corresponding properties of the $\Lambda_b$ baryon
considering the heavy quark symmetry.

The article is arranged as follows:  we study the in-medium properties
  of the heavy baryons $\Lambda_c$ and $\Lambda_b$
 with  the  QCD sum rules in Sec.2; in Sec.3, we present the
numerical results and discussions; and Sec.4 is reserved for our
conclusions.

\section{In-medium properties  of the $\Lambda_Q$ baryons with  QCD sum rules}
We study the  $\Lambda_Q$ baryons  in the nuclear matter
with the two-point correlation functions $\Pi(q)$,
\begin{eqnarray}
\Pi(q) &=& i\int d^4 x\, e^{iq\cdot x}\langle \Psi_0| T\left[J(x)\bar{J}(0)\right]| \Psi_0\rangle \, , \nonumber \\
J(x)&=&\epsilon^{ijk} u^t_i(x)C\gamma_5 d_j(x) Q_k(x)\, ,
\end{eqnarray}
where the $i$, $j$, $k$ are color indexes, $Q=c,b$, and  the $ |\Psi_0\rangle$ is the nuclear matter ground state.

The correlation functions  $\Pi(q)$ can  be decomposed as
\begin{eqnarray}
\Pi(q)&=& \Pi_s(q^2,q \cdot u)+ \Pi_q(q^2,q \cdot u) \!\not\!{q} +\Pi_u(q^2,q \cdot u)  \!\not\!{u}\,,
\end{eqnarray}
according to Lorentz covariance and parity, time reversal invariance.
In  the rest frame of the nuclear matter, $u_\mu=(1,0)$,  $\Pi_i(q^2,q \cdot u)\rightarrow\Pi_i(q_0,\vec{q})$,  $i=s,q,u$.

We can insert  a complete set  of intermediate baryon states with the
same quantum numbers as the current operators $J(x)$ into the correlation functions $\Pi(p)$ to obtain the hadronic representation
\cite{SVZ79}. After isolating the pole terms of the  $\Lambda_Q$ baryons, we use  the dispersion relation to express the
invariant functions  $\Pi_i(q_0,\vec{q})$ in the following form,
\begin{eqnarray}
\Pi_i(q_0,\vec{q})&=&{1\over2\pi i}\int_{-\infty}^\infty~d\omega{\Delta\Pi_i(
\omega,\vec{q})\over\omega-q_0} \, ,
\end{eqnarray}
where
\begin{eqnarray}
\Delta\Pi_s(\omega,\vec{q})&=&-2\pi i \frac{\lambda_{\Lambda_Q}^{*2}M_{\Lambda_Q}^*}{2E^*_q}\left[
\delta(\omega-E_q)-\delta(\omega-\bar{E}_q)\right]\,,\nonumber\\
\Delta\Pi_q(\omega,\vec{q})&=&-2\pi i\frac{\lambda_{\Lambda_Q}^{*2} }{2E^*_q}\left[\delta(\omega-E_q)
-\delta(\omega-\bar{E}_q)\right]\,,
\nonumber\\
\Delta\Pi_u(\omega,\vec{q})&=&+2\pi i\frac{\lambda_{\Lambda_Q}^{*2}\Sigma_v }{2E^*_q}\left[\delta
(\omega-E_q)
-\delta(\omega-\bar{E}_q)\right]\,,
 \end{eqnarray}
 $E_q^*= \sqrt{ M_{\Lambda_Q}^{*2}+\vec{q}^2}$,
$E_q=\Sigma_v+E_q^*$, $\bar{E}_q=\Sigma_v-E_q^*$, and the $M_{\Lambda_Q}^{*}$, $\Sigma_v$ and $\lambda^*_{\Lambda_Q}$ are
 the in-medium masses, vector self-energies and pole residues of the $\Lambda_Q$ baryons, respectively.

At finite nuclear density, the invariant functions $\Pi_i(q_0,\vec{q})$ at the level of quark-gluon degrees of freedom
can be written as \cite{Drukarev1991,C-parameter},
\begin{eqnarray}
\Pi_i(q_0,\vec{q})&=&\sum_n C_n^i(q_0,\vec{q})\langle{O}_n\rangle_{\rho_N}\,,
\end{eqnarray}
where the $C_n^i(q_0,\vec{q})$ are the Wilson coefficients,  the in-medium condensates  $\langle{O}_n\rangle_{\rho_N}=\langle \Psi_0|{O}_n|\Psi_0\rangle =\langle{\cal{O}}\rangle+\rho_N\langle
{\cal{O}}\rangle_N$  at the low nuclear density, the   $\langle{\cal{O}}\rangle$ and $\langle
{\cal{O}}\rangle_N$ denote the vacuum condensates and nuclear matter induced condensates,  respectively. One can consult Refs.\cite{Drukarev1991,C-parameter} for the
technical details in  the operator product expansion.
We carry out the operator product expansion in the nuclear matter  at the large space-like region $q^2\ll 0$ and obtain the QCD  spectral densities, then
take the limit $u_\mu=(1,0)$   and obtain the imaginary parts,
\begin{eqnarray}
\Delta\Pi_i(\omega,\vec{q})&=&{\rm{limit}}_{\epsilon\to 0} \left[\Pi_i(\omega+i\epsilon,\vec{q})-\Pi_i(\omega-i\epsilon,\vec{q}) \right]\,.
\end{eqnarray}

We can match   the phenomenological side with the QCD side of the spectral densities,
and multiply both sides with  the weight function $(\omega-\bar{E}_q)e^{-\frac{\omega^2}{M^2}}$, which
excludes  the  negative-energy pole contribution,  then perform
the integral  $\int_{-\omega_0}^{\omega_0}d\omega$,
\begin{eqnarray}
\int_{-\omega_0}^{\omega_0}d\omega\Delta\Pi_i(\omega,\vec{q})(\omega-\bar{E}_q)e^{-\frac{\omega^2}{M^2}}\,,
\end{eqnarray}
where the $\omega_0$ is the threshold parameter,
finally  obtain the following QCD  sum rules:
\begin{eqnarray}
\lambda_{\Lambda_Q}^{*2}e^{-\frac{E_q^2}{M^2}} &=&\int_{m_Q^2}^{s_0^*}ds\int_{x_i}^1 dx \left\{ \frac{3x(1-x)^2(s-\widehat{E}_Q^2)^2}{128\pi^4}-\frac{(1-x)^2m_Q^2}{384\pi^2x^2} \langle\frac{\alpha_sGG}{\pi}\rangle_{\rho_N}\delta(s-\widehat{E}_Q^2)\right. \nonumber\\
&&+\frac{x}{128\pi^2}\langle\frac{\alpha_sGG}{\pi}\rangle_{\rho_N}-\frac{x\langle q^{\dagger}iD_0q\rangle_{\rho_N}}{6\pi^2}+\frac{x(1-x)\langle q^{\dagger}iD_0q\rangle_{\rho_N}}{3\pi^2}\left[1+2s\delta(s-\widehat{E}_Q^2)\right] \nonumber\\
&&+\bar{E}_q\left[  \frac{x(1-x)\langle q^\dagger q\rangle_{\rho_N}}{4\pi^2}  -\frac{x\langle q^\dagger iD_0iD_0 q\rangle_{\rho_N}}{2\pi^2}\delta(s-\widehat{E}_Q^2)
+\frac{x(1-x)\langle q^\dagger iD_0iD_0 q\rangle_{\rho_N}}{2\pi^2} \right. \nonumber\\
&&\left(1+\frac{2s}{M^2}\right)\delta(s-\widehat{E}_Q^2)-\frac{x\langle q^{\dagger} g_s \sigma G q\rangle_{\rho_N}}{12\pi^2}\delta(s-\widehat{E}_Q^2)
+\frac{x(1-x)\langle q^{\dagger} g_s\sigma G q\rangle_{\rho_N}}{24\pi^2}\left(1+\frac{2s}{M^2}\right)\nonumber\\
&&\left.\left.\delta(s-\widehat{E}_Q^2)
+\frac{x\langle q^\dagger g_s\sigma G q\rangle_{\rho_N}}{16\pi^2}\delta(s-\widehat{E}_Q^2) \right]\right\}e^{-\frac{s}{M^2}}+\frac{\langle\bar{q}q\rangle^2_{\rho_N}+\langle q^\dagger q\rangle^2_{\rho_N}}{6}e^{-\frac{E_Q^2}{M^2}}\, ,
\end{eqnarray}
\begin{eqnarray}
\frac{\lambda_{\Lambda_Q}^{*2}M_{\Lambda_Q}^*}{m_Q}e^{-\frac{E_q^2}{M^2}} &=&\int_{m_Q^2}^{s_0^*}ds\int_{x_i}^1 dx \left\{ \frac{3(1-x)^2(s-\widehat{E}_Q^2)^2}{128\pi^4}-\frac{(1-x)^2s}{384\pi^2x} \langle\frac{\alpha_sGG}{\pi}\rangle_{\rho_N}\delta(s-\widehat{E}_Q^2)\right. \nonumber\\
&&+\frac{(1-x)^3}{192\pi^2x^2}\langle\frac{\alpha_sGG}{\pi}\rangle_{\rho_N}+\frac{1}{128\pi^2}\langle\frac{\alpha_sGG}{\pi}\rangle_{\rho_N}-\frac{\langle q^{\dagger}iD_0q\rangle_{\rho_N}}{6\pi^2}+\frac{(1-x)\langle q^{\dagger}iD_0q\rangle_{\rho_N}}{3\pi^2} \nonumber\\
&&\left[1+2s\delta(s-\widehat{E}_Q^2)\right]+\bar{E}_q\left[  \frac{(1-x)\langle q^\dagger q\rangle_{\rho_N}}{4\pi^2} -\frac{\langle q^\dagger iD_0iD_0 q\rangle_{\rho_N}}{2\pi^2}\delta(s-\widehat{E}_Q^2)
 \right. \nonumber\\
&&+\frac{(1-x)\langle q^\dagger iD_0iD_0 q\rangle_{\rho_N}}{2\pi^2}\left(1+\frac{2s}{M^2}\right)\delta(s-\widehat{E}_Q^2)-\frac{\langle q^{\dagger}g_s \sigma G q\rangle_{\rho_N}}{12\pi^2}\delta(s-\widehat{E}_Q^2)
\nonumber\\
&&\left.\left.+\frac{(1-x)\langle q^{\dagger}g_s \sigma G q\rangle_{\rho_N}}{24\pi^2}\left(1+\frac{2s}{M^2}\right)\delta(s-\widehat{E}_Q^2)
+\frac{\langle q^{\dagger}g_s \sigma G q\rangle_{\rho_N}}{16\pi^2}\delta(s-\widehat{E}_Q^2)
\right]\right\}e^{-\frac{s}{M^2}}\nonumber\\
&&+\frac{\langle\bar{q}q\rangle^2_{\rho_N}+\langle q^\dagger q\rangle^2_{\rho_N}}{6}e^{-\frac{E_Q^2}{M^2}}\, ,
\end{eqnarray}
\begin{eqnarray}
\lambda_{\Lambda_Q}^{*2}\Sigma_v e^{-\frac{E_q^2}{M^2}} &=&\int_{m_Q^2}^{s_0^*}ds\int_{x_i}^1 dx \left\{ \frac{x(1-x)^2(s-\widehat{E}_Q^2)\langle q^{\dagger} q\rangle_{\rho_N}}{8\pi^2}
-\frac{x\langle q^{\dagger}iD_0iD_0q\rangle_{\rho_N}}{4\pi^2}
+\frac{3x(1-x)}{4\pi^2}\right. \nonumber\\
&&\langle q^{\dagger}iD_0iD_0q\rangle_{\rho_N}\left[1+2s\delta(s-\widehat{E}_Q^2)\right]-\frac{x\langle q^{\dagger}g_s\sigma G q\rangle}{24\pi^2}+\frac{x(1-x)\langle q^{\dagger}g_s \sigma G q\rangle_{\rho_N}}{16\pi^2}\left(1+\frac{2s}{M^2}\right) \nonumber\\
&&\left.\left.\delta(s-\widehat{E}_Q^2)+\bar{E}_q\left[  \frac{2x(1-x)\langle q^{\dagger}iD_0 q\rangle_{\rho_N}}{3\pi^2}\right.
-\frac{x\langle q^{\dagger}g_s \sigma G q\rangle_{\rho_N}}{32\pi^2}\delta(s-\widehat{E}_Q^2) \right]\right\}e^{-\frac{s}{M^2}}\, ,
\end{eqnarray}
where $\widehat{E}_Q^2=\frac{m_Q^2}{x}+\vec{q}^2$, $E_Q^2=m_Q^2+\vec{q}^2$, $x_i=\frac{m_Q^2}{s}$, $s^*_0=\omega_0^2=s_0-\vec{q}^2$,
the  $x_i$  in the spectral densities where the function $\delta(s-\widehat{E}_Q^2)$  appears should be $0$. We can obtain the in-medium masses $M^*_{\Lambda_Q}$, vector
self-energies $\Sigma_v$ and pole residues $\lambda^*_{\Lambda_Q}$ by solving above equations with simultaneous  iterations.

\section{Numerical results and discussions}
In calculations, we have assumed that  the linear density
approximation   is valid at the low nuclear  density. The input parameters are taken as
 $\langle q^\dagger q\rangle_{\rho_N}={3\over2}\rho_N$,
 $\langle \bar{q} q\rangle_{\rho_N}=\langle \bar{q} q\rangle+{\sigma_N\over
m_u+m_d}\rho_N $,
$\langle\frac{\alpha_sGG}{\pi}\rangle_{\rho_N}=\langle\frac{\alpha_sGG}{\pi}\rangle-(0.65\pm0.15)\,{\rm GeV}\rho_N$,
$\langle q^\dagger iD_0 q\rangle_{\rho_N}=(0.18\pm0.01)\,{\rm GeV}\rho_N$,
$\langle q^{\dagger} iD_0iD_0 q\rangle_{\rho_N}+{1 \over 12}\langle q^{\dagger}g_s\sigma G q\rangle_{\rho_N}=0.031\,{\rm{GeV}}^2\rho_N$,
$\langle \bar{q} iD_0iD_0 q\rangle_{\rho_N}+{1\over 8}\langle\bar{q}g_s\sigma G q\rangle_{\rho_N}=0.3\,{\rm{GeV}}^2\rho_N$,
$\langle\bar{q}g_s\sigma G q\rangle_{\rho_N}=\langle\bar{q}g_s\sigma G q\rangle+3.0\,{\rm GeV}^2\rho_N$,
$\langle q^{\dagger}g_s\sigma G q\rangle_{\rho_N}=-0.33\,{\rm GeV}^2\rho_N$,
$\langle\bar{q}g_s\sigma Gq\rangle=m_0^2\langle\bar{q}q\rangle$, $\langle\bar{q}q\rangle=-(0.23\pm0.01\,\rm{GeV})^3$, $m_0^2=(0.8\pm0.2)\,\rm{GeV}^2$, $\langle\frac{\alpha_sGG}{\pi}\rangle=(0.33\,\rm{GeV})^4$,
$m_u+m_d=12\,\rm{MeV}$,
$\sigma_N=(45\pm 10)\,\rm{MeV}$, $\vec{q}^2=(0.27\,\rm{GeV})^2$,
$\rho_N=(0.11\,\rm{GeV})^3$, $\langle \bar{q}q\rangle^2_{\rho_N}=f\langle \bar{q}q\rangle_{\rho_N}\times\langle \bar{q}q\rangle_{\rho_N}+(1-f)\langle \bar{q}q\rangle \times \langle \bar{q}q\rangle$, and $f=0.5\pm 0.5$ \cite{C-parameter}.

In the limit $\rho_N=0$, we can recover the QCD sum rules in the vacuum, see Eqs.(8-9). We can take  the
threshold parameters $s^0_{\Lambda_c}=9.6\,\rm{GeV}^2$ and
$s^0_{\Lambda_b}=42.0\,\rm{GeV}^2$  determined in Ref.\cite{Wangzg} to reproduce the experimental data.  Taking the Borel parameters as
$M^2=(2.5-3.5)\,\rm{GeV}^2$ and $(4.3-5.3)\,\rm{GeV}^2$ for the $\Lambda_c$ and $\Lambda_b$ baryons, respectively, we   obtain the
 hadronic parameters $M_{\Lambda_c}=2.284^{+0.049}_{-0.078}\,\rm{GeV}$, $M_{\Lambda_b}=5.618^{+0.078}_{-0.104}\,\rm{GeV}$,
 $\lambda_{\Lambda_c}=(0.022\pm0.002)\,\rm{GeV}^3$,
   $\lambda_{\Lambda_b}=(0.027\pm0.003)\,\rm{GeV}^3$,
the uncertainties originate from the Borel parameters $M^2$.
The central values of the masses are consistent with the experimental data  $M_{\Lambda_c}=2.28646\,\rm{GeV}$ and $M_{\Lambda_b}=5.6202\,\rm{GeV}$ \cite{PDG}.

\begin{figure}
 \centering
 \includegraphics[totalheight=5cm,width=6cm]{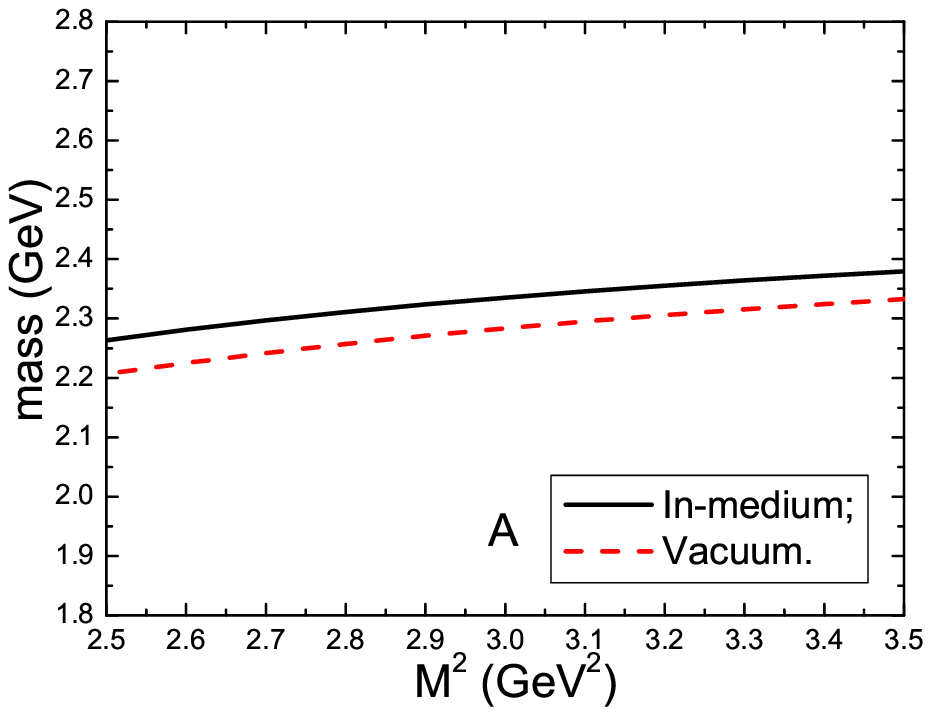}
 \includegraphics[totalheight=5cm,width=6cm]{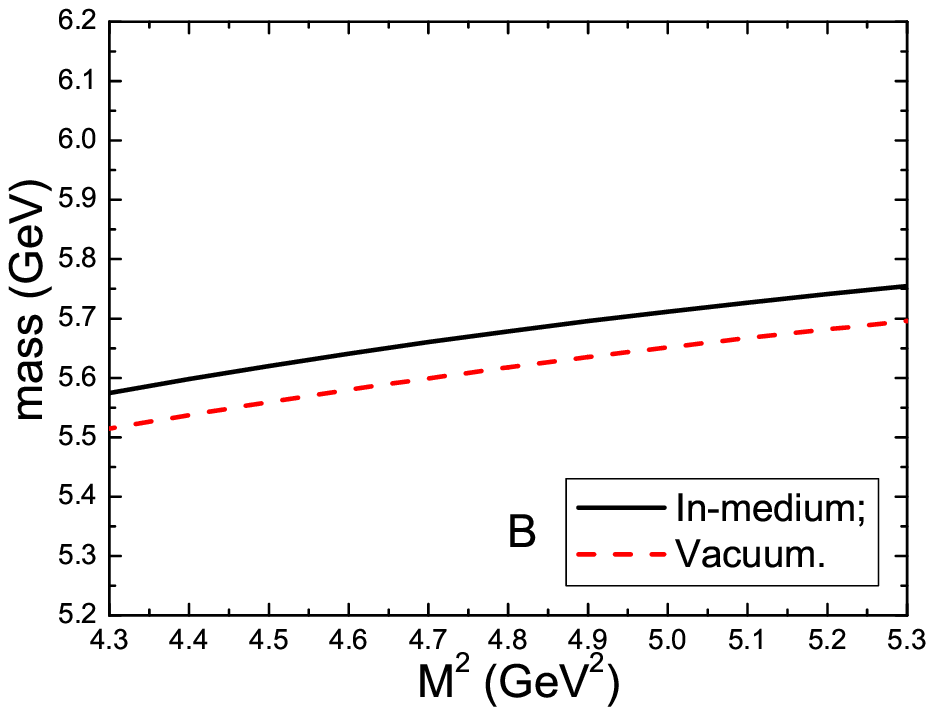}
 \caption{The   in-medium and vacuum masses  versus the Borel parameter $M^2$, the $A$ and $B$ denote the
 $\Lambda_c$ and $\Lambda_b$ baryons, respectively. }
\end{figure}

\begin{figure}
 \centering
 \includegraphics[totalheight=5cm,width=6cm]{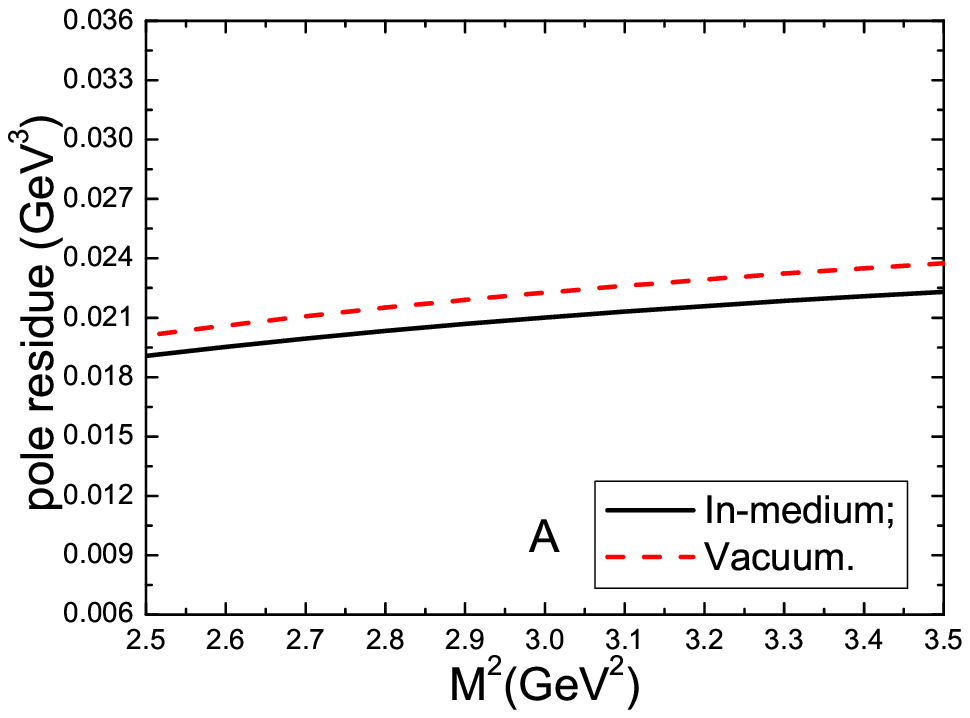}
 \includegraphics[totalheight=5cm,width=6cm]{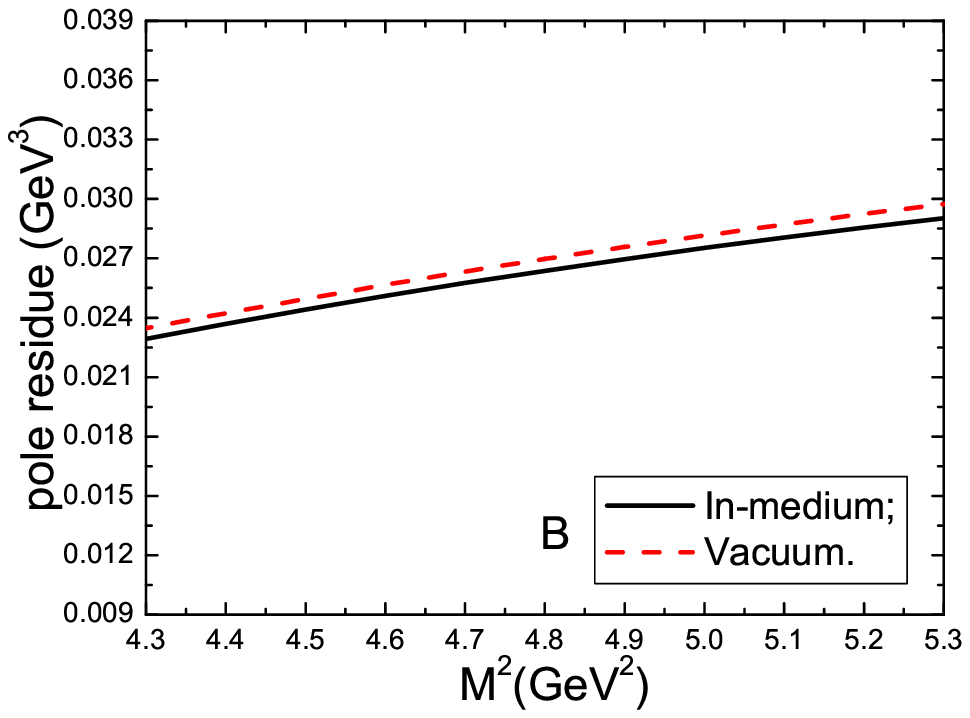}
 \caption{The   in-medium and vacuum pole residues  versus the Borel parameter $M^2$, the $A$ and $B$ denote the
 $\Lambda_c$ and $\Lambda_b$ baryons, respectively. }
\end{figure}

Taking   the same Borel parameters and threshold parameters as  the QCD sum rules in the vacuum, we can obtain the in-medium hadronic parameters
 of the $\Lambda_Q$ baryons,
$M_{\Lambda_c}^*=2.335^{+0.045}_{-0.072}\,\rm{GeV}$,
 $M_{\Lambda_b}^*=5.678^{+0.077}_{-0.103}\,\rm{GeV}$, $\Sigma^{\Lambda_c}_v=(34\pm1)\,\rm{MeV}$,  $\Sigma^{\Lambda_b}_v=(32\pm1)\,\rm{MeV}$, $\lambda_{\Lambda_c}^*=(0.021\pm0.001)\,\rm{GeV}^3$,
 $\lambda_{\Lambda_b}^*=(0.026\pm0.003)\,\rm{GeV}^3$,
 again the uncertainties originate from the Borel parameters. In Figs.1-2, we present the numerical values of the in-medium masses and pole residues versus the Borel parameter $M^2$ compared with the QCD sum rules in the vacuum.
 We can eliminate some uncertainties by introducing the mass differences  $\Delta M_{\Lambda_Q}=M_{\Lambda_Q}^*-M_{\Lambda_Q}$, and obtain
the values  $\Delta M_{\Lambda_c}=51\,\rm{MeV}$ and
 $\Delta M_{\Lambda_b}=60\,\rm{MeV}$.
 The ratios are $\frac{\Delta M_{\Lambda_c}}{M_{\Lambda_c}}=2.2\%$  and $\frac{\Delta M_{\Lambda_b}}{M_{\Lambda_b}}=1.1\%$, respectively,
 the mass modifications are slight, on the other hand, $\frac{\lambda_{\Lambda_Q}-\lambda_{\Lambda_Q^*}}{\lambda_{\Lambda_Q}}\approx 4\%$,
 the pole residue  modifications are also slight,
 we expect that the threshold parameters survive in the nuclear matter, and the in-medium effects cannot
 modify the pole contributions significantly. In calculations, we observe that  the  exponential  factor $e^{-\frac{s_0}{M^2}}\ll e^{-1}$ in the Borel windows,
 the contributions from the higher  resonance and continuum states are greatly suppressed. There are uncertainties for the threshold parameters, in general, we expect
 that $(M_{ gr}+\frac{\Gamma_{ gr}}{2})^2\leq s_0 \leq (M_{ re}-\frac{\Gamma_{ re}}{2})^2$, where the $gr$ and $re$ stand for the ground state and first radial excited state (or resonance), respectively.  At this interval, there are some values which can satisfy the two criterions (pole dominance and convergence of the operator product expansion) of the QCD sum rules. In this article, we take the values $\delta s_0$ determined in our previous work \cite{Wangzg}.
 In Fig.3, we plot the contributions from the perturbative term  and $\langle\bar{q}q\rangle^2_{\rho_N}+\langle q^{\dagger}q\rangle^2_{\rho_N}$ term versus the Borel parameter $M^2$ in the operator
 product expansion. From the figure, we can see that the dominant (or main) contributions come from the perturbative term,  while the contributions of the dimension-6 term $\langle\bar{q}q\rangle^2_{\rho_N}+\langle q^{\dagger}q\rangle^2_{\rho_N}$ are  about $(6-19)\%$  (or $(14-29)\%$) for the $\Lambda_c$ (or $\Lambda_b$) baryon in the Borel window, the operator  product expansion is well convergent. Compared with the corresponding QCD sum rules in the vacuum, the QCD sum rules in the nuclear matter have  better
 convergent behavior in the operator product expansion, see Fig.3.

 \begin{figure}
 \centering
 \includegraphics[totalheight=5cm,width=6cm]{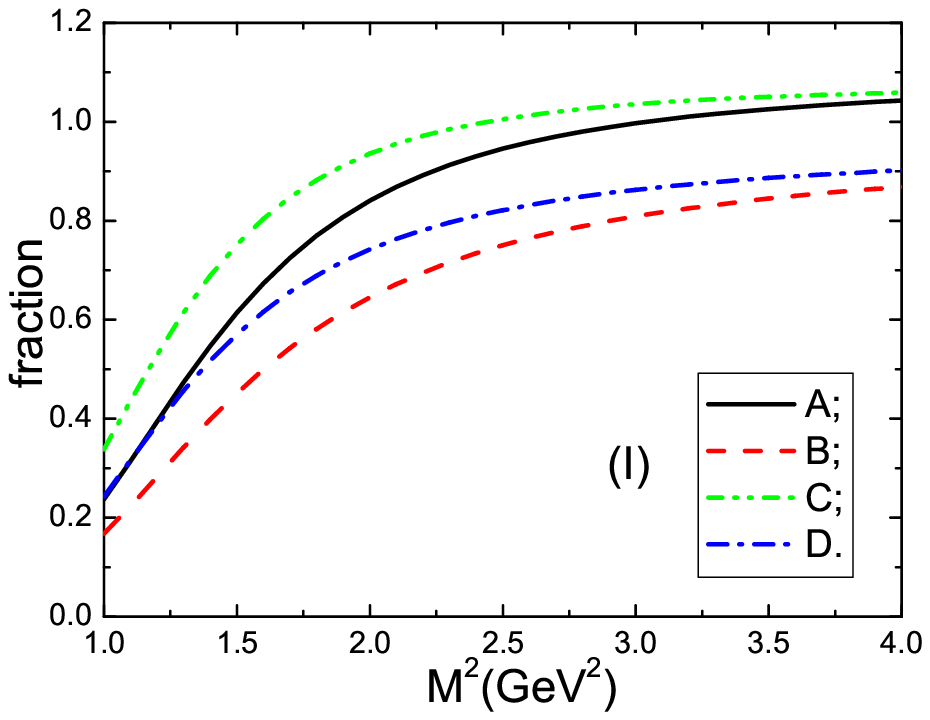}
 \includegraphics[totalheight=5cm,width=6cm]{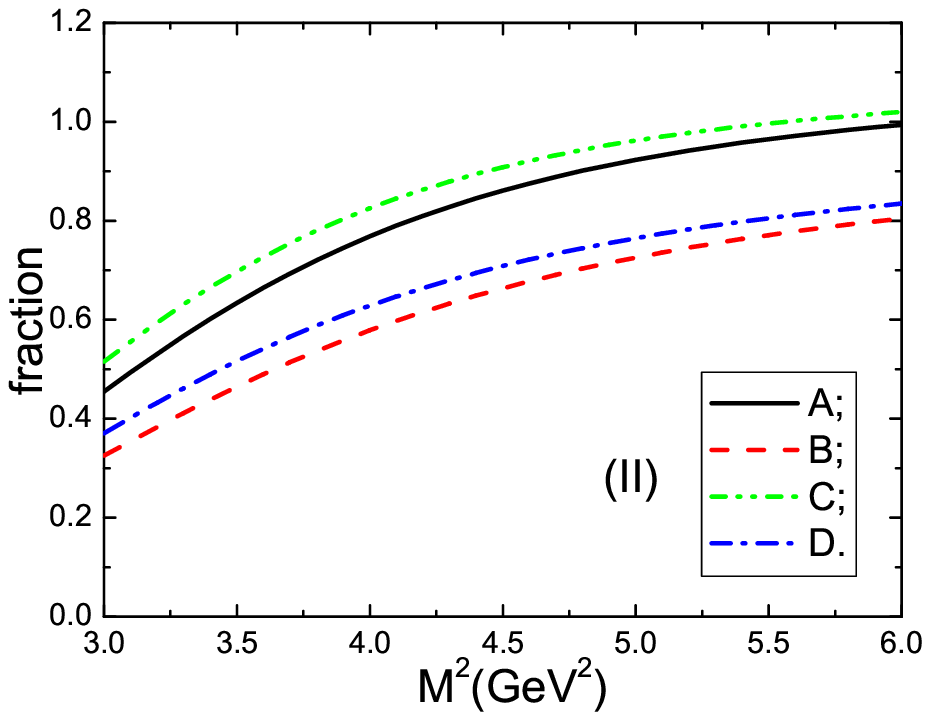}
 \includegraphics[totalheight=5cm,width=6cm]{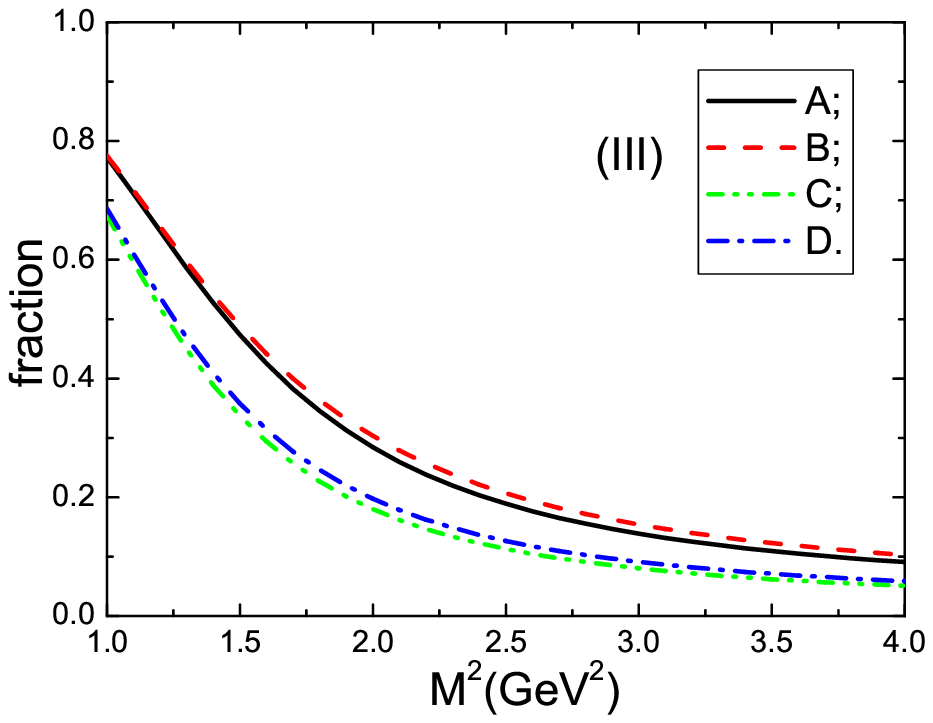}
 \includegraphics[totalheight=5cm,width=6cm]{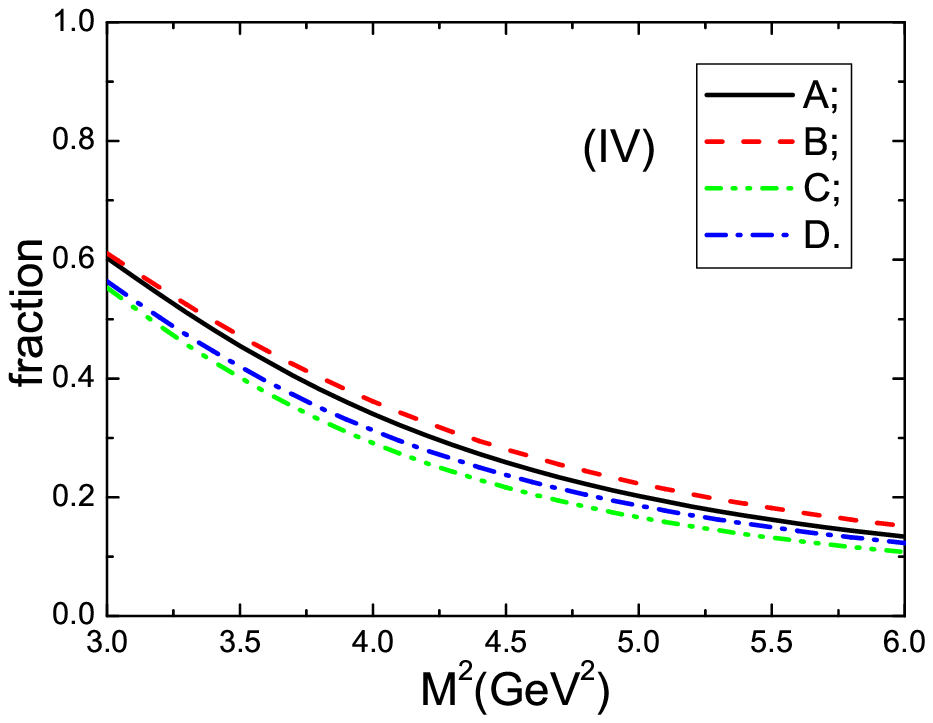}
 \caption{The  contributions from the perturbative term  and $\langle\bar{q}q\rangle^2_{\rho_N}+\langle q^{\dagger}q\rangle^2_{\rho_N}$ term versus the Borel parameter $M^2$ in the operator
 product expansion. (I), (II) and (III), (IV) correspond the perturbative term and $\langle\bar{q}q\rangle^2_{\rho_N}+\langle q^{\dagger}q\rangle^2_{\rho_N}$ term, respectively. (I), (III) and (II), (IV) correspond to the $\Lambda_c$ and $\Lambda_b$ baryons, respectively.  $A$, $B$ and $C$, $D$ correspond to the structures $\!\not\!{q}$ and $1$, respectively. $A$, $C$ and $B$, $D$ correspond to
 the in-medium and in-vacuum QCD sum rules, respectively. }
\end{figure}

    In this article, we have neglected the contributions of the perturbative $\mathcal {O}(\alpha_s)$  corrections, which can be taken into account by introducing the formal coefficient $1+\frac{\alpha_s}{\pi}f(m_Q,s)$ through the unknown function $f(m_Q,s)$. As the  dominant (or main) contributions come from the perturbative term, we expect that the  $\mathcal {O}(\alpha_s)$  corrections to the perturbative term cannot change the predictions greatly. If
   the perturbative $\mathcal {O}(\alpha_s)$  corrections have the typical value $30\%$, i.e. $1+\frac{\alpha_s}{\pi}f(m_Q,s)=1.3$, where we have neglected the mass $m_Q$ and energy $s$ dependence to make a rough estimation, the resulting mass-shifts are $\Delta M_{\Lambda_c}=43\,\rm{MeV}$ and $\Delta M_{\Lambda_b}=59\,\rm{MeV}$. In calculations, we observe that the mass-shifts $\delta M$ decrease with the increase of the perturbative contributions.

The uncertainties originate from the $\delta\langle\frac{\alpha_sGG}{\pi}\rangle_{\rho_N}$,
$\delta\langle q^\dagger iD_0 q\rangle_{\rho_N}$,  $\delta m_0^2$, $\delta\sigma_N$ are tiny and can be  neglected safely.
The uncertainty of the quark condensate $\delta\langle\bar{q}q\rangle$ leads to somewhat larger uncertainties for the masses $M_{\Lambda_Q}^*$ and $M_{\Lambda_Q}$,
but  the mass-shifts $\Delta M$ change slightly and the  uncertainties can be neglected.
 We can take into account the uncertainties of the heavy quark masses  and threshold parameters,  $\delta m_c=0.1\,\rm{GeV}$, $\delta m_b=0.1\,\rm{GeV}$, $\delta s^0_{\Lambda_c}=0.6\,\rm{GeV}^2$,  $\delta s^0_{\Lambda_b}=1.2\,\rm{GeV}^2$ \cite{Wangzg}, as the dominant (or main) contributions come from the perturbative term, those uncertainties maybe result in considerable uncertainties for the mass differences $\Delta M$. In calculations, we observe that the values of the mass differences  $\Delta M_{\Lambda_c}=51\,\rm{MeV}$ and
 $\Delta M_{\Lambda_b}=60\,\rm{MeV}$ survive approximately, while other parameters obtain the uncertainties,  $\delta\lambda_{\Lambda_c}^*=(\pm0.002\pm0.003)\,\rm{GeV}^3$,
 $\delta\lambda_{\Lambda_b}^*=(\pm0.004\pm0.004)\,\rm{GeV}^3$, $\delta\Sigma^{\Lambda_c}_v=(\pm 1\pm 1)\,\rm{MeV}$, and
  $\delta\Sigma^{\Lambda_b}_v=(\pm 1\pm 1)\,\rm{MeV}$.

 At the interval $f=0\sim1$, the masses $M^*_{\Lambda_Q}$ increase monotonously
 with the increase of the parameter $f$, the uncertainty $\delta f=\pm 0.5$   result in the uncertainties
   $\delta M^*_{\Lambda_c}={}^{+68}_{-59}\,\rm{MeV}$ and  $\delta M^*_{\Lambda_b}={}^{+113}_{-90}\,\rm{MeV}$. The corresponding  uncertainties of other hadronic parameters are rather small,
   $|\delta\lambda_{\Lambda_c}^*|<0.001\,\rm{GeV}^3$,  $|\delta\lambda_{\Lambda_b}^*|\leq0.002\,\rm{GeV}^3$, $|\delta\Sigma^{\Lambda_c}_v|\leq 1\,\rm{MeV}$, $|\delta\Sigma^{\Lambda_b}_v|\leq 2\,\rm{MeV}$, and can be neglected.

For the nucleons,  $m_u\approx m_d\approx0$, the in-medium hadronic parameters $ M_N^*$ and $\Sigma_v$ can be approximated as
 $M_N^*=-\frac{8\pi^2}{M^2}\langle\bar{q}q\rangle_{\rho_N}$ and $\Sigma_v=\frac{64\pi^2}{3M^2}\langle q^{\dagger}q\rangle_{\rho_N}$,
respectively, and the in-medium mass modification is  large, about $40\%$. In the present case,
the quark masses $m_c$ and $m_b$ are large, no such simple relations  can be obtained, and the mass modifications  are rather mild.
In the QCD sum rules, the  modifications of the phenomenological spectral densities   are related to the in-medium  quark and gluon condensates. In the linear
density approximation, the in-medium modification of the quark condensate is large while the modification of the gluon condensate is mild,
 $ \langle \bar{q} q\rangle_{\rho_N}\approx0.60 \langle \bar{q} q\rangle$ and $\langle \frac{\alpha_sGG}{\pi}\rangle_{\rho_N}\approx0.93\langle \frac{\alpha_sGG}{\pi}\rangle$ \cite{C-parameter}. Furthermore, there appear additional quark condensates associated with the light flavors, such as
 $\langle q^{\dagger}q\rangle_{\rho_N}$, $\langle q^\dagger iD_0 q\rangle_{\rho_N}$, $\langle q^\dagger iD_0 iD_0q\rangle_{\rho_N}$, $\langle \bar{q}  iD_0 iD_0q\rangle_{\rho_N}$, etc, which also play an important role. The $\Lambda_Q$ baryons have a heavy quark besides two light quarks,
  the heavy quark  interacts  with the nuclear matter through  the exchange of the intermediate gluons, we  expect
 the mass modifications  are smaller than that of the nucleons, which have three light quarks, $uud$ or $udd$.

\section{Conclusion}
In this article, we
  study the in-medium properties  of the
heavy baryons $\Lambda_c$ and $\Lambda_b$ using the QCD sum rules, and obtain the analytical expressions for
 the in-medium masses $M_{\Lambda_Q}^*$,
vector self-energies $\Sigma_v$ and pole residues $\lambda^*_{\Lambda_Q}$, then  get the values   $M_{\Lambda_c}^*=2.335\,\rm{GeV}$,
 $M_{\Lambda_b}^*=5.678\,\rm{GeV}$, $\Sigma^{\Lambda_c}_v=34\,\rm{MeV}$,  $\Sigma^{\Lambda_b}_v=32\,\rm{MeV}$,
 $\lambda_{\Lambda_c}^*=0.021\,\rm{GeV}^3$,
 $\lambda_{\Lambda_b}^*=0.026\,\rm{GeV}^3$,
 $M_{\Lambda_c}^*-M_{\Lambda_c}=51\,\rm{MeV}$,
 $M_{\Lambda_b}^*-M_{\Lambda_b}=60\,\rm{MeV}$,  and discuss the revelent uncertainties.
 The present predictions  can be confronted with the experimental data in the future.

\section*{Acknowledgments}
This  work is supported by National Natural Science Foundation,
Grant Number 11075053,  and the Fundamental
Research Funds for the Central Universities.

\end{document}